\documentclass[aip,jmp,nofootinbib]{revtex4-1}
\usepackage{graphics}
\usepackage{graphicx}
\usepackage{dcolumn}
\usepackage{bm}
\usepackage{mathrsfs}
\usepackage{pstricks}
\usepackage{color}
\usepackage{slashed}
\usepackage{amsmath}
\usepackage{epsfig}
\usepackage{amsfonts}
\usepackage{amssymb}
\def\beq{\begin{equation}}
\def\eeq{\end{equation}}
\def\bea{\begin{eqnarray}}
\def\eea{\end{eqnarray}}

\def\Re{\textrm{Re}}

\begin{document}

\title
{Thermodynamics of the Bosonic \\Randomized Riemann Gas }
%
%
%

\author{J. G. Due\~nas}
\email{jgduenas@cbpf.br}
\affiliation{Centro Brasileiro de Pesquisas F\'{\i}sicas, Rio de Janeiro, RJ 22290-180,
Brazil}

\author{N. F. Svaiter}
\email{nfuxsvai@cbpf.br}
\affiliation{Centro Brasileiro de Pesquisas F\'{\i}sicas, Rio de Janeiro, RJ 22290-180,
Brazil}

\begin{abstract}

The partition function of a bosonic Riemann gas is given by the Riemann zeta function.
We assume that the hamiltonian of this gas at a given temperature $\beta^{-1}$ has a random variable $\omega$ with
a given probability distribution over an ensemble of hamiltonians.
We study the average free energy density and average mean energy density of this arithmetic gas in the complex $\beta$-plane.
Assuming that the ensemble is made by an enumerable infinite set of copies, there is a critical temperature where the average free energy density
diverges due to the pole of the Riemann zeta function.
Considering an ensemble of non-enumerable set of copies, the average
free energy density is non-singular for all temperatures, but acquires complex values in the critical region.
Next, we study the mean energy density of the system which depends strongly on the distribution of the non-trivial zeros of the Riemann zeta function.
Using a regularization procedure we prove that the this quantity is continuous and bounded for finite temperatures.

\end{abstract}

\pacs{02.10.De, 05.30.Jp, 05.40.-a, 11.10-z}

\maketitle

\section{Introduction}

\quad $\,\,$
Arithmetic quantum theory has been developed to establish connections between number theory and
quantum field theory \cite{stn,spec1,bakas,julia,spec2,spec3, wolf}. A bosonic Riemann gas
is a second quantized mechanical system at temperature $\beta^{-1}$,
with partition function given by the Riemann zeta function \cite{riem,ingham}. As was
discussed by Weiss and collaborators,
the hamiltonian for the Riemann gas can in principle be produced in a Bose-Einstein condensate \cite{Weiss}.
The purpose of this article is to study the consequences of introduce randomness in the bosonic Riemann gas.
We study the thermodynamic variables of this arithmetic gas in the complex $\beta$-plane.

In equilibrium statistical mechanics, macroscopic systems can develop phase
transitions when external parameters change. To give a mathematical description of
these phenomena it is usual to define the free energy, $F_{N}(\beta)$, where $N$ is the
number of particles of the system. For finite systems, the free energy
and the partition function are analytic in the entire complex  $\beta$-plane. In the
thermodynamic limit, at points where analyticity is not preserved, one says that a phase transition
occurs. The singularities in the free energy density corresponds to the zeros of the partition function in the complex $\beta$-plane.

In two papers, Lee and Yang \cite{yl1,yl2} studied the zeros of the partition
function of the Ising model in the complex magnetic field. They obtained the following theorem:
in the Ising model in a complex magnetic field $h$, the complex zeros of the partition function are located on
the unit circle in the complex activity plane. Discussing the zeros in the complex temperature plane, Fisher studied a
Ising model on a square two-dimensional
lattice and showed that there is an accumulation of these complex zeros close to the critical point \cite{fisher1}.
Systems with disorder and randomness also have been considered \cite{de1,de2}.
A very simple example of a disordered system which exhibits a phase transition is the random energy model.
The zeros of the partition function in
the complex temperature plane in this model has been studied numerically and also analytically \cite{di}.
In disordered systems, Matsuda, Nishimori and Hukushima studied the distribution of zeros of the partition function in Ising spin
glasses on the complex field plane \cite{matsuda}.
More recently Takahashi and collaborators studied the zeros of
the partition function in the mean-field spin-glass models \cite{takahashi0,takahashi2,takahashi1}.

We discuss the thermodynamics of an arithmetic gas introducing randomness in the system.
Although a probabilistic approach in number theory is not new in the literature \cite{granville, france},
as far as we know the approach presented here is discussed for the first time. We assume that the hamiltonian of the bosonic gas
contains a random variable with some probability distribution. The thermodynamic quantities must be calculated averaging over an ensemble of
realizations of the random quantity. Since the random variable is a parameter in an extensive quantity,
we have to perform ensemble average of some extensive quantity of interest, in our case the free energy \cite{dotsenko, ni}.

Assuming that the ensemble is made by a enumerable infinite set of copies, characterized by the
parameter $\omega_{k}$ defined in the interval $\{\omega:\,\omega_{1}\leq \omega_{k} < \infty \}_{k\in \mathbb{N}}$, the first copy the ensemble with
parameter $\omega_1$ provides a logarithmic singular contribution to the
free energy density due to the pole of the Riemann zeta function.
This temperature, where the average free energy density diverges is the Hagedorn temperature \cite{hagedorn} of the random system.
On the other hand, considering an ensemble of non-enumerable set of copies, the singular behavior of
the average free energy density disappears.
Meanwhile, due to the behavior of the Riemann zeta function in the critical region,
the average free energy density acquires complex values.
Finally, we show that the non-trivial zeros of the Riemann zeta function, which are the Fisher zeros of the system, contribute to
the average energy density of the system.
For other approaches discussing the physics of Riemann zeros see ~\cite{bohigas,for2,livro,rev}. Also, in the literature we can find
other recent results connecting number theory and quantum field theory ~\cite{gn1,gn2,andrade,ds, duenas2}.

Note that it is possible to obtain the Riemann zeta function as the partition function of a prime
membrane \cite{lapidus}.
Let $\mathbb{T}$ be the adelic infinite dimensional torus defined as the adelic product of
circles of lengths $\frac{1}{\ln\,2}$,$\frac{1}{\ln\,3}$,...,$\frac{1}{\ln\,p}$,...,
where $p$ ranges through the sequence of prime numbers. The normalized eigenfrequencies of
${\mathbb{T}}_{p}$ are given by $\nu_{m,p}=m\,\ln\,p$, with $m=0,1,2,...,$.
The spectral partition function of the  ${\mathbb{T}}_{p}$ is $Z_{p}(s)=(1-p)^{-1}$, which
is exactly the factor that appears in the Euler product for the Riemann zeta function.

The organization of the paper is as follows.  In section II we discuss bosonic and fermionic arithmetic gases and discuss also the
singularity structure for the average free energy density of the bosonic Riemann random gas on the complex $\beta$-plane.
In section III the average energy density for the ensemble of Riemann random gas is presented. In section IV we use superzeta functions in order to define the
average energy density of the Riemann random gas. In section V the regularized average energy density is presented. Conclusions are given in section VI.
In this paper we use $k_{B}=c=\hbar=1$.

\section{The partition function of the Riemann gas and the average free energy density of the system}

Let us assume a non-interacting bosonic field theory defined in a volume $V$ with
hamiltonian given by
\begin{equation}
H_{B}=\omega\sum_{k=1}^{\infty}\ln(p_{k})b^{\dagger}_{k}b_{k},
\label{25}
\end{equation}
where $b_k^{\dagger}$ and $b_{k}$ are respectively the creation and annihilation operators and
$\{p_{k}\}_{k\in \mathbb{N}}$ is the sequence of prime numbers. Since the energy of each mode
is $\nu_{k}=\omega\ln\,p_{k}$ the partition function of this system is
exactly the Riemann zeta function, i.e, $Z_{B}=\zeta(\beta\omega)$.

Let us consider now the same situation for a system composed by fermions. The hamiltonian of a free fermionic arithmetic gas is
\begin{equation}
H_F=\omega\sum_{k=1}^{\infty}\ln(p_{k})c^{\dagger}_{k}c_{k},
\label{26}
\end{equation}
where $c_k^{\dagger}$ and $c_{k}$ are respectively the creation and annihilation operators of quanta associated
to the fermionic field and $\{p_{k}\}_{k\in \mathbb{N}}$ is again the sequence of prime numbers.

To proceed, let us introduce the M\"obius function $\mu(n)$ defined by ~\cite{landau},
\begin{displaymath}
\mu(n)=\left\{
\begin{array}{ll}
1,\,\,& \mbox{if n=1,}\\
(-1)^r,\,\,& \mbox{if n is the product of r ($\geq 1$) distinct primes,}\\
0\,\,& \mbox{otherwise, i.e., if the square of at least one prime divides n.}\\
\end{array}\right.
\end{displaymath}\\
Using the M\"obius function it is possible to show that the partition function of the fermionic system is
given by $\zeta(\beta\omega)/\zeta(2\beta\omega)$~{\cite{bakas,spec3}.
As discussed in the literature, from the partition function associated to the hamiltonians $H_{B}$ and $H_{F}$ we get
\begin{equation}
Z_{F}(\beta\omega)\,Z_{B}(2\beta\omega)=Z_{B}(\beta\omega).
\end{equation}
The noninteracting mixture of a bosonic and a fermionic gas with respective temperatures $\beta^{-1}$ and
$(2\beta)^{-1}$ is equivalent to another bosonic gas with
temperature $\beta^{-1}$.

Disorder can now be introduced simply assuming that the parameter $\omega$ that appears in the hamiltonian for the
arithmetic gas given by Eq. (\ref{25}) is a random variable. We take $\{\omega_{k}\}_{k\in\mathbb{N}}$ a set of uncorrelated
random variables with some probability
distribution $P(\omega_{k})$ over an ensemble of hamiltonians. To proceed
we have to perform ensemble averages of some
extensive quantity of interest.
Let us define the average free energy of the Riemann gas as
\begin{equation}
\langle\,F(\beta)\rangle=-\frac{1}{\beta}\langle\,\ln \zeta(\beta\omega)\rangle,
\end{equation}
where $\langle(...)\rangle$ denotes the averaging over an ensemble of realizations of random variable
with a given discrete probability distribution function.
The average free energy density, for the case where the ensemble consists of an enumerable infinite set of copies of the system is
\begin{equation}
\langle\,f(\beta)\rangle=-\frac{1}{\beta\,V}\sum_{k=1}^{\infty}P(\omega_{k})\ln\zeta(\omega_{k}\beta),
\label{free}
\end{equation}
where $V$ is the volume of the system and $P(\omega_{k})$ is a given one-dimensional discrete distribution function
defined in the interval $\{\omega:\,\omega_{1}\leq \omega_{k} < \infty \}_{k\in \mathbb{N}}$.
Note that for low temperatures all of the copies of the ensemble contribute to the average free energy density of the system.
Nevertheless, due to the pole of the Riemann zeta function there is a critical temperature where the first copy of the ensemble gives
a singular contribution to the average free energy density of the system. This is the Hagedorn temperature of the random arithmetic gas.

In the next section, we first show that considering an ensemble made by a  non-enumerable set of copies, the singular behavior of the
average free energy density disappears.
We are also interested in studying the average energy density of the system, which is related to
the logarithmic derivative of the zeta function $\frac{\zeta'}{\zeta}(s)$.
As we will see, the divergent contributions that appear in the average energy density of the system can be circumvented using an
analytic regularization procedure.

\section{The singular structure for the average free energy density and the average energy density for the system}

In this section we will show that the thermodynamic quantities associated
with the Riemann random gas are defined in terms of some number theoretical formulas. With this aim, our next task is
to calculate relevant thermodynamic physical quantities as the average energy density $\langle\,\varepsilon\,\rangle$
and the average entropy density $\langle\,s\,\rangle$.
These thermodynamic quantities are given respectively by
\begin{equation}
\langle\,\varepsilon\,\rangle=-\frac{1}{V}\frac{\partial}{\partial\beta}\sum_{k=1}^{\infty}P(\omega_{k})\ln\zeta(\omega_{k}\beta)
\label{energy}
\end{equation}

and
\begin{equation}
\langle\,s\,\rangle=\frac{1}{V}\biggl(1-\beta\frac{\partial}{\partial\beta}\biggr)\sum_{k=1}^{\infty}P(\omega_{k})\ln\zeta(\omega_{k}\beta).
\label{entropy}
\end{equation}

The properties of the model depend strongly on the analytic structure of the Riemann zeta function.
In the following, instead of considering an ensemble made by a enumerable infinite set of
copies, we extend
these definitions for a non-enumerable set of copies. This approach is analogous to the one that makes the
classical Gibbs ensemble in phase space a continuous fluid \cite{penrose}. Therefore
the average over an ensemble of realizations can be represented by an integral with
$\omega$ defined in the continuum i.e. $\{\omega:\omega\in \mathbb{R}^{+}\}$. The average free energy density and average energy density can be written as
\begin{equation}
\langle\,f(\beta,\lambda)\rangle=-\frac{1}{\beta\,V}\int_{0}^{\infty}\,d\omega\,P(\omega,\lambda)\ln\zeta(\omega\beta)
\label{2}
\end{equation}
and
\begin{equation}
\langle\,\varepsilon(\beta,\lambda)\rangle=-\frac{1}{V}\int_{0}^{\infty}\,d\omega\,P(\omega,\lambda)\frac{\partial}{\partial\beta}\ln\zeta(\omega\beta),
\label{3}
\end{equation}
where
$\lambda$ is a parameter with length dimension that we have to introduce to give the correct dimension in the expressions.
For simplicity let us assume that the probability density distribution is given by
$P(\omega,\lambda)=\lambda e^{-\lambda\omega}$. At this point, a comment may be useful. It is important to stress that the
choice of the $P(\omega)$ does not affect the
conclusions of the paper. By changing the variable $s=\omega\beta$ we can write the average free energy density as

\begin{equation}
\langle\,f(\beta,\lambda)\rangle=-\frac{\lambda}{\beta^{2}V}\int_{0}^{\infty}\,ds\,e^{-\frac{\lambda}{\beta}s}\ln\zeta(s).
\label{4}
\end{equation}
Although there is a singularity in the integrand at $s=1$, it is easy to show that the integral is bounded.
On the neighborhood of $s=1$ we can substitute the function $\log\zeta(s)\approx\log\frac{1}{s-1}=-\log(s-1)$.
This is an integrable singularity, therefore the average free energy density is non-singular and continuous for all temperatures.
Note that overcomes the Hagedorn temperature implies to go into the critical region of the Riemann zeta function, i.e, the region $0 < s < 1$,
that may lead to complex values of the average free energy density.
We would like to point out that
there is an alternative formulation of quantum mechanics and quantum field theory with
non-Hermitian Hamiltonians\,\cite{bender}. The main problem to go into the critical region
is the existence of a branch point of $\ln(z)$ at the origin. This fact generates an ambiguity in the free
energy density. It is clear that this problem disappear if we deal with the logarithmic derivative of the zeta function $\frac{\zeta'}{\zeta}(s)$.
The picture that emerges from this discussion is that the mean energy is a well behaved function of the temperature.

As we discussed above, the logarithmic derivative of the zeta function $\frac{\zeta'}{\zeta}(s)$, which is fundamental in the study of the
density of non-trivial zeros of the zeta function, must be used in the
definition of the average energy density. Let us discuss briefly the symmetries of this set of numbers.
Using the function $\xi(s)$, the functional equation for $\zeta(s)$ given by Eq.~(\ref{f}) takes the form $\xi(s)=\xi(1-s)$. Therefore, if $\rho$ is a zero
of $\xi(s)$, then so is $1-\rho$. Since $\bar{\xi}(\rho)=\xi(\bar{\rho})$ we have that $\bar{\rho}$ and
$1-\bar{\rho}$ are also zeros. The zeros are symmetrically
arranged about the real axis and also about the critical line.
Let us write the complex zeros of the zeta function, which are the Fisher zeros of the system, as $\rho=\frac{1}{2}+i\gamma$, $\gamma \in  \mathbb{C}$. The Riemann hypothesis
is the statement that all $\gamma$ are real. We assume the Riemann hypothesis.
If the zeros $\rho=\frac{1}{2}+i\gamma$, with $\gamma>0$ are arranged in a sequence $\rho_{k}=\frac{1}{2}+i\gamma_{k}$ so that
$\gamma_{k+1}>\gamma_{k}$.

The Riemann zeta function $\zeta(s)$ satisfies the functional equation
\begin{equation}
\pi^{-\frac{s}{2}}\Gamma\biggl(\frac{s}{2}\biggr)\zeta(s)=\pi^{-\frac{(1-s)}{2}}\Gamma\biggl(\frac{1-s}{2}\biggr)\zeta(1-s),
\label{f}
\end{equation}
for $s \in \mathbb{C}\setminus\left\{0,1\right\}$. Let us define the entire function $\xi(s)$ as
\begin{equation}
\xi(s)=\frac{1}{2}s(s-1)\pi^{-\frac{s}{2}}\Gamma\biggl(\frac{s}{2}\biggr)\zeta(s).
\end{equation}
This function $\xi(s)$ has an infinitely many zeros. If they are denoted by $\rho$, Hadamard product for $\xi(s)$ is of the form
\begin{equation}
\xi(s)={e^{b_{0}+b_{1}s}}\,\prod_{\rho}\biggl(1-\frac{s}{\rho}\biggr)\,e^{\frac{s}{\rho}},
\end{equation}
where $b_{0}$ and $b_{1}$ are given constants. Combining both equations we get
\begin{equation}
\frac{1}{2}s(s-1)\pi^{-\frac{s}{2}}\Gamma\biggl(\frac{s}{2}\biggr)\zeta(s)={e^{b_{0}+b_{1}s}}
\,\prod_{\rho}\biggl(1-\frac{s}{\rho}\biggr)\,e^{\frac{s}{\rho}}.
\end{equation}
Taking the logarithmic derivative of both sides of the above equation we get
\begin{eqnarray}
\frac{\zeta'}{\zeta}(s)&&= C_{1}-\frac{1}{s-1}+\sum_{\rho}\frac{1}{(s-\rho)}+\sum_{\rho}\frac{1}{\rho}
+\sum_{n=1}^{\infty}\frac{1}{(s+2n)}-\sum_{n=1}^{\infty}\frac{1}{2n},
\label{1}
\end{eqnarray}
where $C_{1}=-1-\frac{\zeta'(0)}{\zeta(0)} $ is an absolute constant, and $\rho$ is the set of
the nontrivial zeros of the Riemann zeta function \cite{karatsuba}.
Substituting the Eq. (\ref{1}) in Eq. (\ref{3}) we can write the average energy density as
\begin{equation}
\langle\,\varepsilon(\beta,\lambda)\rangle=\varepsilon_{1}(\lambda)+\varepsilon_{2}(\beta,\lambda)+\varepsilon_{3}(\beta,\lambda)+
\varepsilon_{4}(\lambda)+\varepsilon_{5}(\beta,\lambda)+\varepsilon_{6}(\lambda),
\label{enerden}
\end{equation}
where each of these terms are given by
\begin{equation}
\varepsilon_{1}(\lambda)=-\frac{C_{1}}{V}\int_{0}^{\infty}\,d\omega\,\omega P(\omega,\lambda),
\label{enr1}
\end{equation}
\begin{equation}
\varepsilon_{2}(\beta,\lambda)=\frac{1}{V}\int_{0}^{\infty}\,d\omega\,\omega P(\omega,\lambda)\frac{1}{\beta\omega-1},
\label{enr2}
\end{equation}
\begin{equation}
\varepsilon_{3}(\beta,\lambda)=-\frac{1}{V}\int_{0}^{\infty}\,d\omega\,\omega P(\omega,\lambda)\sum_{\rho}\frac{1}{\beta\omega-\rho},
\label{enr3}
\end{equation}
\begin{equation}
\varepsilon_{4}(\lambda)=-\frac{1}{V}\int_{0}^{\infty}\,d\omega\,\omega P(\omega,\lambda)\sum_{\rho}\frac{1}{\rho},
\label{enr4}
\end{equation}
\begin{equation}
\varepsilon_5(\beta,\lambda)=-\frac{1}{V}\int_{0}^{\infty}\,d\omega\,\omega P(\omega,\lambda)\sum_{n=1}^{\infty}\frac{1}{(\beta\omega+2n)}
\label{enr5}
\end{equation}
and finally
\begin{equation}
\varepsilon_6(\lambda)=\frac{1}{2V}\int_{0}^{\infty}\,d\omega\,\omega P(\omega,\lambda)\sum_{n=1}^{\infty}\frac{1}{n}.
\label{enr6}
\end{equation}
Usually, the information of the thermodynamics of the system is contained in derivatives of the mean free energy density.
With the exception of the  contribution coming from the pole of the zeta function, the above
expressions for the mean free energy density are divergent series.
We can use a standard regularization procedure to give meaning to these divergent terms ~\cite{hardy}. Here,
we choose to use an analytic regularization procedure introduced in quantum field theory in ~\cite{bollini} and used extensively since then.

\section{The superzeta function and the average energy density for the arithmetic random gas}

In this section we will discuss each of the terms that contributes to the average energy density given by Eq. (\ref{enerden}).
The contribution of the first term given by Eq. (\ref{enr1}) to the average energy density can be written as
\begin{equation}
\varepsilon_1(\lambda)= -\frac{C_{1}}{\lambda V}.
\end{equation}
The second term that contributes to the average energy density $\varepsilon_2(\beta)$ can be written as
\begin{equation}
\varepsilon_2(\beta,\lambda)= \frac{1}{\beta V}-\frac{\lambda}{\beta^2 V}e^{-\frac{\lambda}{\beta}}Ei(\lambda/\beta),
\end{equation}
where $Ei(x)$ is the exponential-integral function \cite{abram, grads} defined by
\begin{eqnarray}
Ei(x)=-\int_{-x}^{\infty}dt\,\frac{e^{-t}}{t}\,\,\,\,\,\, x<0,
\end{eqnarray}
and
\begin{eqnarray}
Ei(x) = -\lim_{\varepsilon\rightarrow 0}\int_{-x}^{-\varepsilon}\,dt\,\frac{e^{-t}}{t} +\int_{-\varepsilon}^{\infty}\,dt\,\frac{e^{-t}}{t} \,\,\,\,\, x>0.
\end{eqnarray}
For the third term that contributes to the average energy density $\varepsilon_3(\beta,\lambda)$ we have
\begin{equation}
\varepsilon_3(\beta,\lambda)= -\frac{\lambda}{V}\int_{0}^{\infty}d\omega\,\omega e^{-\lambda\omega}\sum_{\rho}\frac{1}{\beta\omega -\rho}.
\end{equation}
As we will see, the contributions given by $\varepsilon_3(\beta,\lambda)$ and $\varepsilon_4(\lambda)$ can be written in terms of superzeta functions.
A straightforward calculation give us for $\varepsilon_4(\lambda)$
\begin{equation}
\varepsilon_4(\lambda)= -\frac{1}{\lambda V}\sum_{\rho}\frac{1}{\rho}.
\end{equation}
Let us discuss the construction of the so-called superzeta or secondary zeta function built over the Riemann zeros, i.e.,
the nontrivial zeros of the Riemann zeta function~\cite{superzeta}.
As was discussed by Voros, in view of the central symmetry of the
Riemann zeros $\rho\leftrightarrow1-\rho$, leads us to generalized zeta functions of several kinds.
Each one of these superzeta functions reflects our choice of set of
numbers to built zeta functions over the Riemann zeros.
The first family that we call $G_{1}(s,t)$ is defined by
\begin{equation}
G_{1}(s,t)=\sum_{\rho}\frac{1}{(\frac{1}{2}+t-\rho)^{s}}\,\,\,\,\,\,\,\,\,\,\, \Re(s)>1,
\end{equation}
valid for $t\,\in \Omega_{1}=\{t\,\in \mathbb{C}\, |\, (\frac{1}{2}+t-\rho)\not\in \mathbb{R}_{-} (\forall\rho)\}$.
This is the simplest generalized zeta-function over the Riemann zeros. The sum runs over all zeros symmetrically and $t$ is just a shift parameter.
In view of the fact that
\begin{equation}
\sum_{\rho}\frac{1}{\rho}=\sum_{k=1}^{\infty}\frac{1}{\frac{1}{2}+i\tau_{k}}+\frac{1}{\frac{1}{2}-i\tau_{k}}=\sum_{k=1}^{\infty}\frac{1}{\frac{1}{4}+\tau_{k}^{2}},
\end{equation}
the second generalized superzeta function is defined as
\begin{equation}
G_{2}(\sigma,t)=\sum_{k=1}^{\infty}(\tau_{k}^{2}+t^{2})^{-\sigma}\,\,\,\,\,\,\,\,\,\,\, \Re(\sigma)>\frac{1}{2},
\label{G2}
\end{equation}
valid for $t\in\, \Omega_{2}=\{t\in\,\mathbb{C}\,|\, t\pm\,i\tau_{k}\not\in\,\pm\,i\mathbb{R}_{-}\, (\forall k)\}.$
The central symmetry $\tau_{k}\leftrightarrow\,-\tau_{k}$,
is preserved in the family of superzeta functions $G_{2}(\sigma,t)$.
Using the above definitions, the contributions to the average energy density given by $\varepsilon_{3}(\beta,\lambda)$ and
$\varepsilon_{4}(\lambda)$ can be written respectively as
\begin{equation}
\varepsilon_{3}(\beta,\lambda)=-\frac{\lambda}{V}\int_{0}^{\infty}\,d\omega\,\omega e^{-\lambda\omega}G_{1}(1,\beta\omega-1/2)
\end{equation}
and
\begin{equation}
\varepsilon_{4}(\lambda)=-\frac{1}{\lambda V}G_{2}(1,1/2).
\end{equation}
Since we are interested in the region $\Re(\sigma)>\frac{1}{2}$ we are in the region of convergence of the series in Eq. (\ref{G2}).
Therefore, assuming the Riemann hypothesis we can write $\varepsilon_{4}(\lambda)$ as
\begin{equation}
\varepsilon_{4}(\lambda)=-\frac{1}{\lambda V}\sum_{n=1}^{\infty}\frac{1}{\frac{1}{4}+\gamma_n^{2}}.
\end{equation}
After this discussion of the terms that involve superzeta functions, let us proceed with the contribution to the average energy density given by terms
that involve the Hurwitz zeta function and the Riemann zeta function.
The Hurwitz zeta function $\zeta(z,q)$ is the analytic extension of the series
\begin{equation}
\zeta(z,q)=\sum_{n=0}^{\infty}\frac{1}{(q+n)^{z}}\,\,\,\,\,\,\,\,\, q\neq 0,-1,-2,..., \,\,\, Re(z)>1,
\label{hurwitz}
\end{equation}
which is a meromorphic function in the whole complex plane with a single pole at $z=1$.
We can write $\varepsilon_{5}(\beta,\lambda)$ in terms of the Hurwitz zeta function as
\begin{equation}
\varepsilon_{5}(\beta,\lambda)=\lim_{z\rightarrow 1}\biggl(\frac{1}{\beta V}-\frac{\lambda}{2 V}
\int_{0}^{\infty}\,d\omega\,\omega e^{-\lambda\omega}\zeta(z,\beta\omega/2)\biggr).
\end{equation}
Finally, let us discuss the last term given by $\varepsilon_{6}(\lambda)$. Using the zeta function $\zeta(s)$ defined by
\begin{equation}
\zeta(s)=\sum_{n=1}^{\infty}\frac{1}{n^{s}}\,\,\,\,\,\,\,\,\,\,\, Re(s)>1,
\end{equation}
and otherwise by its analytic extension, we can write $\varepsilon_{6}(\beta,\lambda)$ as
\begin{equation}
\varepsilon_{6}(\lambda)=\lim_{s\rightarrow 1}\frac{1}{2\lambda V}\zeta(s).
\end{equation}

In the next section we will discuss the structure of the singularities for $\varepsilon_{3}(\beta,\lambda)$,
$\varepsilon_{5}(\beta,\lambda)$ and $\varepsilon_{6}(\lambda)$ to the average energy density.

\section{The regularized average energy density}

The aim of this section is to find the contributions given by $\varepsilon_{3}(\beta,\lambda)$,
$\varepsilon_{5}(\beta,\lambda)$ and $\varepsilon_{6}(\lambda)$ to the average energy density. From the previous section we
have that the contribution to the average energy density given by $\varepsilon_{3}(\beta,\lambda)$ can be written as
\begin{equation}
\varepsilon_{3}(\beta,\lambda)=-\frac{\lambda}{V}\int_{0}^{\infty}\,d\omega\,\omega e^{-\lambda\omega}G_{1}(1,\beta\omega-1/2).
\end{equation}
The superzeta function $G_{1}(s,t)$ admits an analytic extension to the half-complex plane given by
\begin{equation}
G_{1}(s,t)=-Z(s,t)+\frac{1}{(t-\frac{1}{2})^s}+\frac{\sin \pi s}{\pi}\mathcal{J}(s,t),
\end{equation}
where $Z(s,t)$ is the superzeta function $G_{1}(s,t)$ evaluated in the trivial zeroes of the Riemann zeta function \cite{superzeta}.
It is possible to write $Z(s,t)$ in terms of the Hurwitz zeta function, defined in Eq. \eqref{hurwitz}, as
\begin{equation}
Z(s,t)=\sum_{k=1}^{\infty}\biggl(\frac{1}{2}+t+2k\biggr)^{-s}=2^{-s}\zeta\biggl(s,\frac{1}{4}+\frac{1}{2}t\biggr).
\end{equation}
The function $\mathcal{J}(s,t)$, a Mellin transform of the logaritmic derivative of the Riemann zeta function, is defined as
\begin{equation}
\mathcal{J}(s,t)=\int_{0}^{\infty}\frac{\zeta'}{\zeta}\biggl(\frac{1}{2}+t+y\biggr)y^{-s}dy\,\,\,\,\,\,\,\, Re(s)<1.
\end{equation}
As discussed by Voros, all the discontinuities of $(t-1/2)^{-s}$ and $-Z(s,t)$ in the interval $(-\infty,\frac{1}{2}]$
cancel against jumps of $\mathcal{J}(s,t)$. Therefore, the analytic continuation of $G_{1}(s,t)$ is a regular function
of $t\in \Omega_1$.
Using the analytical properties of the Mellin transformation of $(\zeta'/\zeta)(x)$ the function $\mathcal{J}(s,t)$ has a
known global meromorphic structure which implies the meromorphic continuation to the whole plane of the superzeta
function $G_1(s,t)$.
The function $\mathcal{J}(s,t)$ has simple poles at $s=+1,+2,...$ with residues given by
$Res(\mathcal{J}(s,t))|_{s=n}=-\frac{1}{(n-1)!}\frac{d^n}{dt^n}\log|\zeta\bigl(\frac{1}{2}+t\bigr)|$, and this structure makes
the product $\sin(\pi s)\, \mathcal{J}(s,t)$ free of singularities. Hence, the singularity structure of the function
$G_1(s,t)$ is the same as the one of $Z(s,t)$, i.e, of the Hurwitz zeta function.

The analytic extension of the Hurwitz zeta function $\zeta(s,q)$ can be performed as given in \cite{karatsuba}.
The result is a meromorphic function for $Re(s)>0$, with a simple pole at $s=1$ and of residue $1$.
It is possible to show that
\begin{equation}
\lim_{s\rightarrow 1}\biggl(\zeta(s,q)- \frac{1}{s-1}\biggr)=-\psi(q),
\label{PVhurwitz}
\end{equation}
where  the $\psi(q)$ is the Euler's Psi-function defined by
\begin{equation}
\psi(x)=\frac{d}{d x}\ln{\Gamma(x)}.
\label{psi}
\end{equation}
Since $\sin(\pi s)$ vanishes at integers, the contribution to the average energy density given by
$\epsilon_3(\beta, \lambda)$ can be written as
\begin{equation}
\epsilon_3(\beta, \lambda)=-\frac{\lambda}{V}\int_{0}^{\infty}d\omega\, \omega e^{-\lambda\omega}
\biggl(\frac{1}{2}\psi\biggl(\frac{\beta\omega}{2}\biggr)+\frac{1}{\beta\omega - 1/4}\biggr).
\end{equation}
To perform the integral of the first term in the parenthesis,  we can use the functional relation for the
psi function $\psi(x+1)=\psi(x)+1/x$ and the following series representation for $\psi(x+1)$
\begin{equation}
\psi(x+1)=-C+\sum_{k=2}^{\infty}(-1)^{k}\zeta(k)x^{k-1}.
\label{psi-relation}
\end{equation}
The integral of the second term is similar to the one calculated for $\epsilon_2(\beta,\lambda)$ which can be expressed in
terms of the exponential-integral function $Ei(x)$. Accordingly, the contribution to the average energy density given by
$\epsilon_3(\beta, \lambda)$ can be written as
\begin{equation}
\epsilon_3(\beta, \lambda)= \frac{C}{2\lambda V}-
\frac{1}{\beta V}\sum_{k=2}^{\infty}g(k)\biggl(\frac{\beta}{\lambda}\biggr)^k+
\frac{\lambda}{4\beta^2 V}e^{-\frac{\lambda}{4\beta}}Ei\biggl(\frac{\lambda}{4\beta}\biggr)
\end{equation}	
where $g(k)=\frac{(-1)^{k}}{2^{k}}\Gamma(k+1)\zeta(k)$.

From the last section, the contribution to the average energy density given by $\varepsilon_{5}(\beta,\lambda)$  can be written as
\begin{equation}
\varepsilon_{5}(\beta,\lambda)=\lim_{s\rightarrow 1}\biggl(\frac{1}{\beta V}-\frac{\lambda}{2 V}\int_{0}^{\infty}\,d\omega\,\omega e^{-\lambda\omega}\zeta(s,\beta\omega/2)\biggr).\nonumber\\
\end{equation}
A similar procedure as the mentioned above in the analysis of the Hurwitz zeta function can be performed with the aid of equations
(\ref{PVhurwitz}), (\ref{psi}) and (\ref{psi-relation}). In this case we can see that $\varepsilon_{5}(\beta,\lambda)$ has three contributions,
the first two are finite and the last one is singular. We have
\begin{equation}
\varepsilon_{5}(\beta,\lambda)=\frac{1}{\beta\,V} + \frac{\lambda}{2 V}\int_{0}^{\infty}\,d\omega\,\omega\,
e^{-\lambda\omega}\psi\biggl(\frac{\omega\beta}{2}\biggl)-h(\lambda,V).
\end{equation}
We can write $\varepsilon_{5}(\beta,\lambda)$ as
\begin{eqnarray}
\varepsilon_{5}(\beta,\lambda)=-\frac{C}{2\lambda V}+\frac{1}{\beta V}\sum_{k=2}^{\infty}g(k)\biggl(\frac{\beta}{\lambda}\biggr)^{k}-h(\lambda,V),
\end{eqnarray}
where $h(\lambda,V)=\lim_{s\rightarrow 1}\bigl(\frac{1}{2\lambda V}\frac{1}{s-1}\bigr)$ and $g(k)$ being the same coefficient
as before. From the previous section we have that the contributions to
the average energy density given by $\varepsilon_{6}(\lambda)$ is
\begin{equation}
\varepsilon_{6}(\lambda)=\lim_{s\rightarrow 1}\frac{1}{2\lambda V}\zeta(s).\nonumber\\
\end{equation}
The $\varepsilon_{6}(\lambda)$ contribution is proportional to the Riemann zeta function.
Using the analytic extension of the Riemann zeta function and the fact that
\begin{equation}
\lim_{s\rightarrow 1}\biggl(\zeta(s)- \frac{1}{s-1}\biggr)=-\psi(1),
\end{equation}
where $\psi(1)=-0.577215$ is the Euler's constant, we can write $\varepsilon_{6}(\beta)$ as a
finite contribution and again a singular part. We have
\begin{equation}
\varepsilon_{6}(\lambda)=-\frac{1}{2\lambda  V}\psi(1)+h(\lambda,V).
\end{equation}
Note that the singular contribution in $\varepsilon_{5}(\beta,\lambda)$ and $\varepsilon_{6}(\beta,\lambda)$
cancel each other.

We can split the contribution of each term to the average energy density into two categories. The first one coming from
$\varepsilon_{A}(\lambda)=\varepsilon_{1}(\lambda)+\varepsilon_{4}(\lambda)+\varepsilon_{6}(\lambda)$. These terms are temperature
independent and, therefore, we can interpret them as coming from the
vacuum modes associated to the arithmetic gas. The other terms $\varepsilon_{2}(\beta ,\lambda)$,
$\varepsilon_{3}(\beta, \lambda)$ and $\varepsilon_{5}(\beta,\lambda)$ are temperature dependent contributions.
We can write this thermal contribution $\epsilon_B(\beta ,\lambda)$ as
\begin{eqnarray}
\varepsilon_{B}(\beta ,\lambda)=&&\frac{1}{\beta V}
-\frac{\lambda}{\beta^2 V}e^{-\frac{\lambda}{\beta}}Ei\biggl(\frac{\lambda}{\beta}\biggr)
+\frac{\lambda}{4 \beta^2 V}e^{-\frac{\lambda}{4\beta}}Ei\biggl(\frac{\lambda}{4\beta}\biggr).
\end{eqnarray}

We have shown that the thermodynamic quantities associated to the arithmetic gas can be calculated.
A similar calculation can be performed to find the average entropy density. Since we have an ambiguity in the mean free energy density the
mean entropy also keeps such ambiguity.

\section{Conclusions}

"Do you know  a physical reason that the Riemann hypothesis should be true?"(E. Landau).
 Hilbert and P\'olya suggested that there might be a spectral interpretation of the the non-trivial zeros of the Riemann
zeta function. The corresponding operator must be self-adjoint. The existence of such operator may led to the proof of the Riemann hypothesis.
In this paper we present a different scenario where it is possible to present some links between the Riemann zeta function theory and physics.
We show that using an arithmetic quantum field theory with randomness it is possible to connect strongly the non-trivial zeros of the Riemann
zeta function with a measurable physical quantity defined in the system.

Arithmetic quantum field theories establish a bridge between statistical mechanic systems with an
exponential density of states and multiplicative number theory.
In these theories, the partition function of  hypothetical systems are related to the Riemann zeta function
or other Dirichlet series. The partition
function of a bosonic Riemann gas is given by $Z_{\beta}=\zeta(\beta\omega)$, where $\beta$ is the inverse of the
temperature of the gas.
Since the Riemann zeta function has a simple pole in $s=1$, there is a Hagedorn
temperature above which the system can not be heated up.
In a bosonic Riemann gas with randomness, we show that the mean energy density depends strongly on the distribution of the Riemann zeros.
Being more
precise, the mean energy density of the system is defined in terms of
the quantity $\frac{\zeta'}{\zeta}(s)$, which is related to the non-trivial zeros of the Riemann zeta function.

First, assuming that the ensemble is made by a enumerable infinite set of copies,
there is a critical temperature where the average free energy density diverges due to the pole of the Riemann zeta function.
This is the Hagedorn phenomenon in this random system.
On the other hand, considering an ensemble of made by a non-enumerable set of copies, the singular behavior of
the average free energy density disappears,
i.e,  it is non-singular for all temperatures. Due to the behavior of the Riemann zeta function in the critical region,
the average free energy density acquires complex values and appears an ambiguity in the free
energy density. Second, we study the average energy density of the system, where the above mentioned problem disappears, since
is related to
the logarithmic derivative of the zeta function $\frac{\zeta'}{\zeta}(s)$.
We showed that the divergent contributions that appear in the average energy density of the system can be circumvented
using an analytic regularization procedure.


The parafermion arithmetic gas of order $r$ is a quantum gas where the
exclusion principle states that no more than $r-1$ parafermions can have the same quantum numbers \cite{green}.
The partition function for an arithmetic parafermion gas or order $r$ is given by $Z_{r}(\beta,\omega)=\zeta(\beta\omega)/\zeta(r\beta\omega)$.
A natural extension of this paper
is to assume that, as the previous case, $\omega$ is a random variable with
a given probability distribution over an ensemble of hamiltonians.

\section{Acknowlegements}

This paper was supported by Conselho Nacional de Desenvolvimento
Cientifico e Tecnol{\'o}gico do Brazil (CNPq). We would like to thanks E. F. M. Curado, F. D. Nobre and B. F. Svaiter for useful discussions.

\end{document}